\documentclass[sigconf]{acmart}
\AtBeginDocument{%
  }

\usepackage{multirow}
\usepackage{amsthm}
\usepackage{booktabs,multirow}
\theoremstyle{remark}

\copyrightyear{2026}
\acmYear{2026}
\setcopyright{cc}
\setcctype{by}
\acmConference[SIGIR '26]{Proceedings of the 49th International ACM SIGIR Conference on Research and Development in Information Retrieval}{July 20--24, 2026}{Melbourne, VIC, Australia}
\acmBooktitle{Proceedings of the 49th International ACM SIGIR Conference on Research and Development in Information Retrieval (SIGIR '26), July 20--24, 2026, Melbourne, VIC, Australia}
\acmDOI{10.1145/3805712.3809616}
\acmISBN{979-8-4007-2599-9/2026/07}

\begin{document}

\title{TimeMM: Time-as-Operator Spectral Filtering for Dynamic Multimodal Recommendation}


\author{Wei Yang}
\authornote{These authors contributed equally to this work.}
\affiliation{%
  \institution{Xiaohongshu Inc.}
  \city{Beijing}
  \country{China}
}
\email{yangwei103@xiaohongshu.com}

\author{Rui Zhong}
\authornotemark[1]
\affiliation{%
  \institution{Xiaohongshu Inc.}
  \city{Beijing}
  \country{China}
}
\email{chenxiao5@xiaohongshu.com}

\author{Zihan Lin}
\affiliation{%
  \institution{Xiaohongshu Inc.}
  \city{Beijing}
  \country{China}
}
\email{linzihan2@xiaohongshu.com}

\author{Xiaodan Wang}
\affiliation{%
  \institution{Xiaohongshu Inc.}
  \city{Beijing}
  \country{China}
}
\email{wangxiaodan@xiaohongshu.com}

\author{Cheng Chen}
\affiliation{%
  \institution{Xiaohongshu Inc.}
  \city{Beijing}
  \country{China}
}
\email{mengde@xiaohongshu.com}

\author{Huan Ren}
\affiliation{%
  \institution{Xiaohongshu Inc.}
  \city{Beijing}
  \country{China}
}
\email{renhuan@xiaohongshu.com}

\author{Yao Hu}
\affiliation{%
  \institution{Xiaohongshu Inc.}
  \city{Beijing}
  \country{China}
}
\email{xiahou@xiaohongshu.com}

\renewcommand{\shortauthors}{Wei Yang et al.}

\begin{abstract}
Multimodal recommendation improves user modeling by integrating collaborative signals with heterogeneous item content. In real applications, user interests evolve over time and exhibit nonstationary dynamics, where different preference factors change at different rates. This challenge is amplified in multimodal settings because visual and textual cues can dominate decisions under different temporal regimes. Despite strong progress, most multimodal recommenders still rely on static interaction graphs or coarse temporal heuristics, which limits their ability to model continuous preference evolution with fine-grained temporal adaptation. To address these limitations, we propose \textbf{TimeMM}, a time-conditioned spectral filtering framework for dynamic multimodal recommendation. TimeMM instantiates \textbf{Time-as-Operator} by mapping interaction recency to a family of parametric temporal kernels that reweight edges on the user--item graph, producing component-specific representations without explicit eigendecomposition. To capture non-stationary interests, we introduce \textbf{Adaptive Spectral Filtering} that mixes the operator bank according to temporal context, yielding prediction-specific effective spectral responses. To account for modality-specific temporal sensitivity, we further propose \textbf{Spectral-Aware Modality Routing} that calibrates visual and textual contributions conditioned on the same temporal context. Finally, a ranking-space \textbf{Spectral Diversity Regularization} encourages complementary expert behaviors and prevents filter-bank collapse. Extensive experiments on real-world benchmarks demonstrate that TimeMM consistently outperforms state-of-the-art multimodal recommenders while maintaining linear-time scalability. The code is available at \url{https://github.com/llm-ml/TimeMM}.
\end{abstract}






\maketitle
%%%%%%%%%%%%%%%%%%%%%%%%%%%%%%%%%%%%%%%%%%%%%%%%%%%%
%%%%%%%%%%%%%%%%%%%%%%%%%%%%%%%%%%%%%%
\section{Introduction}
\label{sec:intro}
Multimodal recommendation improves collaborative filtering by incorporating rich item side information, such as visual content and textual descriptions, to alleviate interaction sparsity and strengthen preference understanding~\cite{wei2019mmgcn,valencia2026vlm4rec,li2025dpu,kan2026conflating}. By integrating collaborative signals with heterogeneous content features, it can represent not only what users interacted with, but also why an item matches their interests~\cite{deldjoo2021content,han2022modality,yang2023multimodal}. In real applications, however, this matching process is inherently time-dependent~\cite{fan2021continuous,li2020time,ye2020time}. Users often maintain stable semantic interests over long horizons, while their visual tastes can shift over shorter horizons~\cite{zhang2025hierarchical,ye2025harnessing}. For example, a user may repeatedly purchase running shoes over an extended period, yet the preferred visual style can evolve from minimalist designs to high-contrast colorways as trends change. These observations suggest that effective multimodal recommendation requires modeling the dynamic evolution of user preferences~\cite{chen2022time,qin2024learning}.

Despite the evident need for temporal modeling, prevailing multimodal approaches largely overlook the time dimension~\cite{jiang2024diffmm,karra2024interarec,zhao2025hierarchical}.
State-of-the-art methods such as AlignRec~\cite{liu2024alignrec}, MMIL~\cite{yang2024multimodal}, and SLMRec~\cite{tao2022self} typically conceptualize user--item interactions as a static bipartite graph.
In this view, historical interactions are aggregated indiscriminately, compressing a user's behavioral timeline into a flat structural snapshot.
This temporal agnosticism is problematic. An interaction from years ago is treated as equally informative as one from yesterday, making it difficult to distinguish obsolete interests from emerging intents~\cite{vaswani2021multimodal,wu2025aligning}.
Consequently, static models often struggle with concept drift.
They tend to favor signals from a user’s global history, which can misalign with the user’s potential visually driven short-term intent~\cite{bian2023multi,lu2024online}.
Bridging this gap calls for a mechanism that injects temporal dynamics into the graph propagation operator itself, rather than appending time as an auxiliary feature, which remains largely underexplored in multimodal recommendation.

However, recognizing the need for temporal dynamics is only the first step.
The key challenge lies in capturing the multi-scale granularity of preference shifts~\cite{huang2022multi,huang2023multi,zhang2025m2rec}.
User decisions are rarely governed by a single temporal scope. Instead, they reflect a superposition of long-term habits and short-term impulses, spanning stable category loyalty and fleeting responses to viral visual trends.
The effective scope varies across users and can change for the same user over time~\cite{jiang2020aspect,xia2022multi}.
Moreover, temporal sensitivity is modality-dependent: visual preferences often react rapidly to aesthetics, whereas semantic interests tend to evolve more steadily~\cite{liu2023dynamic,zhang2025m2rec}.
This naturally suggests a spectral perspective on graph signals, where different temporal scopes correspond to different smoothing strengths and thus different frequency emphases.

To capture such dynamics, a promising direction is to examine the spectral properties of user behaviors~\cite{xu2025fagcl,zhang2025m2rec,yang2025structured}.
High-frequency signals often correspond to transient and impulse-driven behaviors, whereas low-frequency components reflect stable and enduring habits.
Prior studies such as FITMM~\cite{yang2025fitmm} and SMORE~\cite{ong2025spectrum} introduced spectral graph modeling into this domain.
By explicitly decomposing representations into orthogonal frequency bands through spectral transformation or eigendecomposition, these methods can separate informative signals from modality-specific noise~\cite{fan2024hierarchical,huang2025learning,kim2025diff}.
However, explicit spectral approaches face a key limitation in dynamic settings.
They typically rely on a global decomposition of a static graph snapshot, yielding fixed spectral bases and filtering behaviors that cannot adapt to time-varying spectral shifts~\cite{kim2025graph,liu2024selfgnn,luo2024spectral}.
As a result, it becomes difficult to determine whether a high-frequency component reflects noise or a genuinely emerging short-term interest.
In addition, the required matrix factorization introduces substantial computational overhead, which limits scalability.
Finally, existing methods often apply uniform filtering across modalities, even though visual signals can be more sensitive to short-range fluctuations than textual semantics.

To bridge static spectral modeling with non-stationary user behavior, we propose \textbf{TimeMM}, a time-conditioned spectral filtering framework for dynamic multimodal recommendation. Our core insight follows the duality between time and frequency: rather than performing costly explicit decomposition, we realize frequency-aware modeling implicitly via \textbf{Time-as-Operator}. Specifically, TimeMM maps interaction recency to a family of parametric temporal kernels that reweight edges on the user--item graph, inducing a lightweight spectral filter bank with distinct smoothing profiles. Short-range kernels emphasize rapidly changing signals, whereas long-range kernels retain stable preference evidence. We then perform \textbf{Adaptive Spectral Filtering} by mixing the operator bank according to user and item temporal context to obtain a prediction-specific effective spectral response. Next, \textbf{Spectral-Aware Modality Routing} calibrates visual and textual contributions conditioned on the temporal context, enabling modality mixtures that are consistent with the active spectral regime. Finally, a ranking-space \textbf{Spectral Diversity Regularization} encourages complementary expert behaviors and prevents filter-bank collapse.

In summary, our main contributions are as follows:
\begin{itemize}
    \item We propose \textbf{Time-as-Operator}, which converts interaction recency into a temporal-kernel-weighted operator bank, enabling efficient time-conditioned spectral filtering on the interaction graph without explicit eigendecomposition.
    \item We develop \textbf{TimeMM}, combining Adaptive Spectral Filtering with Spectral-Aware Modality Routing to calibrate visual and textual contributions, and a ranking-space Spectral Diversity Regularization to prevent expert collapse.
    \item Extensive experiments on real-world benchmarks show that TimeMM consistently outperforms strong multimodal baselines while maintaining linear-time scalability.
\end{itemize}

\section{Related Work}
\label{sec:related_work}

\subsection{General Multimodal Recommendation}
Multi-modal contents are widely used to complement interaction data and alleviate sparsity in collaborative filtering~\cite{kang2017visually,yang2023based,li2023biased,li2024panoptic,li2024domain,li2026recgoat}. Many methods augment interaction-based ranking with multimodal features to learn content-aware preferences~\cite{xu2023musenet,yu2022graph,han2022modality,li2024simcen,li2025revisiting}. With graph learning, recommendation is commonly formulated on the user--item interaction graph via topology-aware message passing. MMGCN~\cite{wei2019mmgcn} and GRCN~\cite{wei2020graph} integrate multimodal signals into propagation, while DualGNN~\cite{wang2021dualgnn} models modality-specific structures for improved fusion. Subsequent work improves robustness and finer integration, including higher-order relation modeling in LATTICE~\cite{zhang2021mining} and denoised relations in FREEDOM~\cite{zhou2023tale}. Self-supervised and contrastive objectives are also adopted to reduce modality noise and strengthen invariance, such as MMGCL~\cite{yi2022multi} and SLMRec~\cite{tao2022self}. Recent advances further enhance cross-modal coherence through intention-aware designs like MMIL~\cite{yang2024multimodal} and alignment-based objectives like AlignRec~\cite{liu2024alignrec}, while PromptMM~\cite{wei2024promptmm} and foundation-model-driven recommenders extend multimodal understanding and reasoning~\cite{gu2025r4ec,liu2024rec,xia2025trackrec,xia2025hierarchical}. Despite strong progress, most approaches still operate on static graphs or only coarsely incorporate time, limiting their ability to model evolving preferences.

Large language models (LLMs) have recently received broad attention~\cite{touvron2023llama,chen2026self,yang2025toward,chang2025survey,ping2026poet}. Their strong reasoning, instruction-following, and generalization abilities have led to successful applications in natural language processing and multimodal understanding~\cite{ye2024domain,li2025personalized,ping2025hdlcore,li2025climatellm,chen2025tourrank}. In recommendation, this trend has motivated generative recommendations. Early works such as TALLRec~\cite{bao2023tallrec} and LlamaRec~\cite{yue2023llamarec} adapt LLMs to recommendation through instruction tuning and recommendation-oriented alignment. Later studies improve item grounding and inference efficiency, such as IDGenRec, Bi-Tokenizer, and other ranking-oriented designs~\cite{tan2024idgenrec,bao2025bi,zhang2024notellm2}. More recently, agentic LLMs~\cite{ping2025verimoa,yang2025learning,yang2025maestro} have attracted growing interest because they support autonomous decision-making, iterative feedback modeling, and memory-enhanced reasoning. Representative examples include Agent4Rec~\cite{zhang2024generative}, AgentCF++~\cite{liu2025agentcf++}, and memory-augmented frameworks such as MemRec and AMEM4Rec~\cite{chen2026memrec,nguyen2026amem4rec}.

\subsection{Spectral Multimodal Recommendation}
Frequency-domain modeling offers a complementary view where propagation can be interpreted as spectral filtering that preserves smooth preference components while suppressing noise~\cite{fan2024hierarchical,luo2024dual,huang2025learning,kim2025diff}. In sequential recommendation, spectral transformations and frequency-aware objectives capture periodic patterns and improve robustness under temporal drift~\cite{zhang2023contrastive,du2023frequency,yang2025farec,zhang2025frequency}, including Fourier-based contrastive views in CFIT4SRec~\cite{zhang2023contrastive} and multi-frequency modules in FAGCL~\cite{xu2025fagcl} and M2Rec~\cite{zhang2025m2rec}. For graph recommendation, polynomial and spectral operator formulations have been explored to enhance expressivity and denoising~\cite{kim2025graph,liu2024selfgnn,luo2024spectral}. In multimodal settings, SMORE~\cite{ong2025spectrum} performs band-wise decomposition and fusion, and robustness-oriented designs can be viewed as implicitly encouraging spectral diversity via augmentation or relation denoising~\cite{zhou2023tale,yi2022multi}. More recently, FITMM~\cite{yang2025fitmm} motivates frequency-wise separate-then-fuse fusion from an information-theoretic perspective, while SSR~\cite{yang2025structured} proposes structured spectral reasoning that modulates band reliability and models cross-band interactions for robust fusion. However, most spectral approaches operate on static graphs and thus inherit a stationarity assumption, making it difficult to reflect time-varying spectral shifts as preferences evolve. %Our work departs from static spectral modeling by modulating the propagation operator with temporal signals, enabling time-conditioned multi-scale filtering without explicit eigendecomposition.

%%%%%%%%%%%%%%%%%%%%%%%%%%%%%%%%%%%%%%%%%%%%%%%%%%%%%%%%%%%%%%%%%%%%%%
%\section{PRELIMINARIES}

%In this work, we consider a recommendation scenario where a set of users $\mathcal{U}$ interact with a set of items $\mathcal{I}$. The user-item interactions are represented by an interaction matrix $Y \in \mathbb{R}^{|\mathcal{U}| \times |\mathcal{I}|}$, where each element $y_{ui} = 1$ indicates that user $u \in \mathcal{U}$ has interacted with item $i \in \mathcal{I}$. Beyond the interaction data, each user and item is associated with a unique ID-based representation, which captures collaborative filtering signals. In addition, each item $i$ is accompanied by multimodal content features, providing additional descriptive information. Specifically, we consider \textit{visual} and \textit{textual} modalities, denoted as $\mathcal{M} = \{v, t\}$, where $v$ and $t$ correspond to the visual and textual representations, respectively. The multimodal representation of an item can be expressed as $\mathbf{X}_i^m \in \mathbb{R}^{d_m}$, where $m \in \mathcal{M}$ and $d_m$ denotes the dimensionality of the modality representation.

%%%%%%%%%%%%%%%%%%%%%%%%%%%%%%%%%%%%%%%%%%%%%%%%%%%%%%%%%%%%%%%%%%%%%%
\section{Preliminaries}%# and Problem Formulation}
\label{sec:preliminaries}

Let $\mathcal{U}=\{u\}_{u=1}^{U}$ and $\mathcal{I}=\{i\}_{i=1}^{I}$ denote the user set and the item set.
We represent the training data as a collection of timestamped interactions
$\mathcal{E}=\{(u,i,t)\mid u\in\mathcal{U},\, i\in\mathcal{I},\, t\in\mathbb{R}^{+}\}$,
where each triplet indicates that user $u$ interacted with item $i$ at time $t$.
We construct a bipartite graph over the node set $\mathcal{V}=\mathcal{U}\cup\mathcal{I}$, where $|\mathcal{V}|=N=U+I$. Each item $i\in\mathcal{I}$ is associated with features from $M$ modalities.
We denote the modality set by $\mathcal{M}$.
For each modality $m\in\mathcal{M}$, we use $\mathbf{F}_{m}\in\mathbb{R}^{I\times d_{m}}$ to denote the raw item feature matrix.
Since modalities may lie in different feature spaces, we apply a modality-specific encoder to project $\mathbf{F}_{m}$ into a shared embedding space of dimension $D$.
We denote the resulting initial node embeddings by $\mathbf{X}^{(0)}_{m}\in\mathbb{R}^{N\times D}$.
For users, $\mathbf{X}^{(0)}_{m}$ is initialized as learnable embeddings.
For items, $\mathbf{X}^{(0)}_{m}$ is initialized with the encoded multimodal features.

\section{Method}
\label{sec:method}

\begin{figure*}
  \centering
  \includegraphics[width=\linewidth]{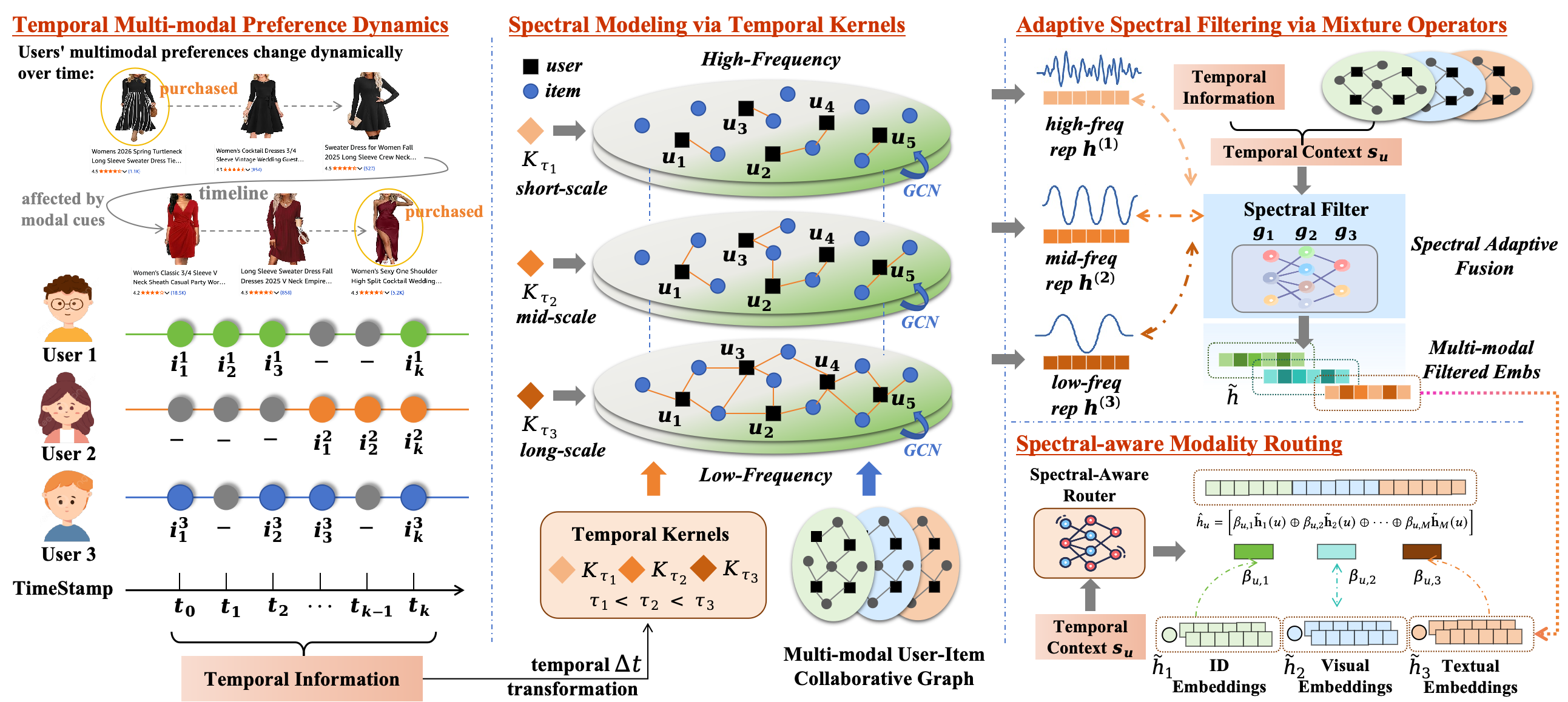}
  \caption{Overall architecture of \textbf{TimeMM}. TimeMM instantiates Time-as-Operator by mapping interaction recency to a family of temporal-kernel-weighted graph operators, forming a lightweight spectral filter bank over the interaction graph. It then performs time-conditioned Adaptive Spectral Filtering by mixing scale-specific propagation outputs according to temporal context, followed by spectral-aware modality routing to calibrate visual and textual contributions under the active spectral regime. A spectral diversity regularizer further encourages complementary expert behaviors and prevents filter-bank collapse.}
  \label{fig:overall_model}
\end{figure*}

In this section, we present the detailed components of TimeMM. Fig.~\ref{fig:overall_model} provides an overview of the overall architecture.

%We propose \textbf{TimeMM}, a time-conditioned spectral filtering framework for multimodal recommendation on the user--item interaction graph. Our core principle is \textbf{Time-as-Operator}: interaction recency is mapped to scale-parameterized temporal kernels that reweight edges and yield a family of normalized propagation operators with distinct smoothing profiles, enabling spectral-style filtering without explicit eigendecomposition. TimeMM then performs \textbf{Adaptive Spectral Filtering} via context-conditioned operator mixing, and applies \textbf{Spectral-aware Modality Routing} to rebalance visual and textual contributions under the selected temporal regime.
%Figure~\ref{fig:overall_model} illustrates the overall architecture.

%%%%%%%%%

%\subsubsection{Learning Objective}
%Given historical interactions with timestamps $\mathcal{E}$ and multimodal features $\{\mathbf{F}_{m}\}_{m\in\mathcal{M}}$, we aim to estimate the preference score of user $u$ over item $i$ for future interactions. A key requirement is to model time-varying preference evolution on the interaction graph, where the effective smoothing strength (and thus the spectral emphasis) may change across users, items, and decision moments. Accordingly, our goal is to learn time-conditioned representations that support adaptive spectral filtering and modality-aware preference scoring.

\subsection{Time-as-Operator: Temporal Kernels as a Spectral Filter Bank}
\label{sec:temporal_diffusion}

A central design in TimeMM is to inject temporal dynamics into the graph operator itself, rather than treating time as an auxiliary side feature.
Starting from the user-item interaction topology, we map continuous timestamps to scale-parameterized edge weights, which induces a family of time-conditioned normalized adjacency operators.
Viewed through the lens of graph signal processing, this operator family behaves as a lightweight \textbf{spectral filter bank}: different temporal scales correspond to different smoothing strengths and thus different frequency responses, enabling multi-scale message passing over the same sparse graph without explicit eigendecomposition.

\subsubsection{Temporal Proximity Kernel}
To characterize how historical interactions contribute to the current preference, we first define the chronological latency of each user--item edge.
Let $t_{\text{anchor}}(u)=\max\{t \mid (u,\cdot,t)\in\mathcal{E}\}$ denote the user-specific temporal anchor (the most recent activity time of user $u$).
For an interaction $(u,i,t_{ui})\in\mathcal{E}$, we measure its recency by
\begin{equation}
\Delta t_{ui}=\max\bigl(0,\, t_{\mathrm{anchor}}(u)-t_{ui}\bigr).
\end{equation}
We posit that temporal influence decays in a heavy-tailed manner, which is commonly observed in human memory and repeated-consumption processes~\cite{wu2007novelty,koren2009collaborative,rizoiu2017expecting}.
Accordingly, we introduce a parametric \textbf{Temporal Proximity Kernel} $\mathcal{K}_{\tau}(\cdot)$ with temporal scale $\tau$:
\begin{equation}
\label{eq:kernel}
w_{ui}^{(\tau)}=\mathcal{K}_{\tau}(\Delta t_{ui})
=\exp\left(-\frac{\log(1+\Delta t_{ui})}{\tau}\right),
\end{equation}
where $\tau$ is a hyper-parameter that controls the effective temporal range.
This kernel provides a soft, differentiable mapping from latency to structural affinity.
A smaller $\tau$ yields faster decay and emphasizes recent interactions, resulting in a weaker smoothing operator, whereas a larger $\tau$ preserves longer history and produces stronger smoothing over broader neighborhoods.

\subsubsection{Spectrum of Propagation Scales}
A single temporal range is often insufficient to capture the superposition of short-term impulses and long-term tastes.
We therefore construct a \textbf{Spectrum of Propagation Scales} by defining a set of scales $\mathcal{T}=\{\tau_1,\tau_2,\ldots,\tau_K\}$.
For each $\tau_k\in\mathcal{T}$, we build a scale-specific weighted adjacency matrix $\mathbf{A}^{(k)}\in\mathbb{R}^{N\times N}$.
For each observed interaction $(u,i)$, we set $\mathbf{A}^{(k)}_{ui}=\mathbf{A}^{(k)}_{iu}=w_{ui}^{(\tau_k)}$, and set $\mathbf{A}^{(k)}_{nj}=0$ for all other node pairs $(n,j)$.

This construction yields $K$ parallel operator views over an identical sparse topology:
\begin{itemize}
    \item \textbf{Short-scale view ($\tau_{\mathrm{small}}$):} prioritizes recent signals, which can better reflect transient or fast-changing behaviors (e.g., reacting to newly emerging visual trends).
    \item \textbf{Long-scale view ($\tau_{\mathrm{large}}$):} retains long-range history, which better captures stable and enduring preferences (e.g., persistent interest in item categories).
\end{itemize}
Crucially, the resulting filter bank is computationally lightweight.
Because all scales share the same sparsity pattern, we only maintain $K$ sets of edge weights and avoid the explicit eigendecomposition or SVD required by classical spectral filtering methods.

\subsubsection{Parallel Multi-scale Propagation}
Given the scale-specific operators, we perform message propagation in parallel for each scale.
We adopt the LightGCN propagation rule~\cite{he2020lightgcn} adapted to weighted graphs.
For scale $k$ and modality $m$, the layer-wise update is
\begin{equation}
\mathbf{X}^{(k,l+1)}_{m}
=\bigl(\mathbf{D}^{(k)}\bigr)^{-1/2}\mathbf{A}^{(k)}\bigl(\mathbf{D}^{(k)}\bigr)^{-1/2}\mathbf{X}^{(k,l)}_{m},
\end{equation}
where $\mathbf{D}^{(k)}$ is the diagonal degree matrix with $\mathbf{D}^{(k)}_{nn}=\sum_{j}\mathbf{A}^{(k)}_{nj}$.
The learnable parameters are shared across scales through the initial embeddings $\mathbf{X}^{(0)}_{m}$, while the scale-specific operators modulate the diffusion behavior.

After $L$ layers, we form the scale-specific representation for node $n$ as
\begin{equation}
\mathbf{h}^{(k)}_{m}(n)=\sum_{l=0}^{L}\mathbf{X}^{(k,l)}_{m}(n).
\end{equation}
This produces $K$ multi-scale representations $\{\mathbf{h}^{(1)}_{m},\ldots,\mathbf{h}^{(K)}_{m}\}$ for each modality, where each scale corresponds to a distinct temporal scope of propagation, and equivalently a distinct spectral smoothing profile.

%\noindent\textbf{Bridge to adaptive filtering.} The construction above defines a fixed \emph{filter bank} indexed by $\tau_k$. However, different users (and different temporal contexts) may require different mixtures of short- vs.\ long-range propagation, and different modalities may exhibit distinct sensitivity to these frequency profiles. In the next section, we therefore introduce a time-conditioned gating mechanism that learns to \emph{adaptively combine} the scale-specific outputs, yielding context-dependent spectral filtering.

\subsection{Adaptive Spectral Filtering via Mixture over Operators}
\label{sec:scale_gating}

The time-parameterized operators in Section~\ref{sec:temporal_diffusion} define a \textbf{spectral filter bank} that produces scale-specific representations $\{\mathbf{h}^{(k)}_m\}_{k=1}^{K}$ for each modality.
However, a single fixed frequency profile is rarely optimal across users, items, and decision moments.
Some preferences are dominated by short-range signals (less smoothing, more high-frequency variation), whereas others depend on long-range history (stronger smoothing, more low-frequency structure).
We therefore introduce \textbf{Adaptive Spectral Filtering}, which performs context-conditioned selection and combination over the operator bank.
Concretely, we view the $K$ scales as a set of experts, and learn a time-conditioned mixture weight that adapts the effective spectral response to each user or item.

\subsubsection{Temporal Context State}
We construct compact temporal state vectors for both users and items as control signals for operator mixing.
For a user $u$, we compute $\mathbf{s}_u\in\mathbb{R}^{d_s}$ to summarize the temporal pattern of the user interaction history.
For an item $i$, we compute $\mathbf{s}_i\in\mathbb{R}^{d_s}$ to summarize the temporal exposure pattern of the item.
These states are derived from training interactions only:
\begin{equation}
\mathbf{s}_u=\phi_u\bigl(\{t \mid (u,\cdot,t)\in\mathcal{E}\}\bigr),
\qquad
\mathbf{s}_i=\phi_i\bigl(\{t \mid (\cdot,i,t)\in\mathcal{E}\}\bigr).
\end{equation}

In our implementation, $\phi_u$ includes statistics such as the log-transformed mean interaction age, the log-transformed median interaction age, the log-transformed activity span, and the fraction of interactions within a recent time window.
Similarly, $\phi_i$ includes statistics such as the item age measured from a global reference time, the item activity span, the interaction count, and the mean time gap between consecutive interactions.
We standardize these features to improve optimization stability.

\subsubsection{Scale-Aware Fusion}
Given the temporal states, we use a learnable gating network to map $\mathbf{s}_u$ and $\mathbf{s}_i$ into \emph{operator-mixture} distributions.
For each user $u$, we compute
\begin{equation}
\mathbf{g}_u=\mathrm{Softmax}\left(\frac{\mathbf{W}_g\,\mathrm{MLP}(\mathbf{s}_u)+\mathbf{b}_g}{T}\right),
\end{equation}
where $\mathbf{g}_u\in\mathbb{R}^{K}$ is the scale allocation over the $K$ operators.
We compute the item distribution $\mathbf{g}_i$ analogously by replacing $\mathbf{s}_u$ with $\mathbf{s}_i$.
$T$ is a temperature hyperparameter controlling the sharpness of the mixture:
smaller $T$ encourages more selective assignment, while larger $T$ produces smoother combinations.
From the spectral perspective, this gating realizes a context-dependent filter by mixing the responses of the operator bank.

We then fuse the scale-specific representations by weighting the $K$ propagation outputs.
For each modality $m$, the fused representation is
\begin{equation}
\tilde{\mathbf{h}}_m(u)=\sum_{k=1}^{K} g_{u,k}\, \mathbf{h}^{(k)}_m(u),
\qquad
\tilde{\mathbf{h}}_m(i)=\sum_{k=1}^{K} g_{i,k}\, \mathbf{h}^{(k)}_m(i).
\end{equation}
This design yields user--item-specific effective temporal scopes, since both sides adapt their spectral profiles via $\mathbf{g}_u$ and $\mathbf{g}_i$.
It also preserves the interpretation of scales as complementary experts, where each expert corresponds to a distinct smoothing regime and thus a distinct frequency emphasis.

%\noindent\textbf{Bridge to modality routing.} While the operator mixture adapts \emph{which} frequency profiles to emphasize, different modalities can still exhibit heterogeneous sensitivity to those profiles. In the next section, we introduce time-conditioned modality routing to adaptively balance visual and textual signals conditioned on the same temporal context.

%%%%%%%%%%%
\subsection{Spectral-Aware Modality Routing}
\label{sec:modality_routing}

The adaptive operator mixture in Section~\ref{sec:scale_gating} determines which spectral profiles (short-range vs.\ long-range smoothing regimes) should be emphasized for each user or item, yielding the fused representations $\tilde{\mathbf{h}}_m(\cdot)$ per modality.
Yet modalities are not equally reliable under the same regime: visual cues may better explain fast-changing engagement patterns, while textual semantics often provide more stable preference signals.
Therefore, after selecting an effective spectral response, we further route modality contributions conditioned on the temporal context.
We introduce \textbf{Spectral-Aware Modality Routing}, which learns a personalized modality mixture consistent with the time-conditioned filtering stage, avoiding brittle hand-assigned modality--scale associations.

\subsubsection{Context-Conditioned Modality Scores}
Let $\tilde{\mathbf{h}}_m(u)\in\mathbb{R}^{D}$ denote the scale-fused user representation for modality $m$ after adaptive spectral filtering.
We map the shared temporal context $\mathbf{s}_u$ to modality scores via a lightweight context encoder:
\begin{equation}
\mathbf{z}_u=\mathrm{MLP}_{\mathrm{mod}}(\mathbf{s}_u),
\qquad
\boldsymbol{\beta}_u=\mathrm{Softmax}(\mathbf{z}_u),
\end{equation}
where $\mathbf{z}_u\in\mathbb{R}^{M}$ and $\boldsymbol{\beta}_u\in\mathbb{R}^{M}$ with $\sum_{m=1}^{M}\beta_{u,m}=1$.
Rather than enforcing representation-level alignment across modalities, $\boldsymbol{\beta}_u$ implements context-conditioned routing that rebalances modality contributions under the spectral regime implied by $\mathbf{s}_u$.

\subsubsection{Routed Multimodal User Representation}
We form the final user representation by re-scaling each modality channel and concatenating them:
\begin{equation}
\hat{h}_u=
\left[
\beta_{u,1}\tilde{\mathbf{h}}_1(u)
\oplus
\beta_{u,2}\tilde{\mathbf{h}}_2(u)
\oplus
\cdots
\oplus
\beta_{u,M}\tilde{\mathbf{h}}_M(u)
\right].
\end{equation}
This routing mechanism supports spectral-consistent modality shifting: short-range regimes can emphasize modalities that capture rapid changes, while long-range regimes can prioritize modalities that provide robust slowly varying preference cues.

%As a result, TimeMM couples time-conditioned spectral filtering with modality routing in a single coherent pipeline.

%\noindent\textbf{Bridge to spectral diversity.} While the above modules enable adaptive filtering and modality routing, the scale experts may still collapse to redundant behaviors. In the next section, we introduce a decorrelation objective to encourage complementary spectral roles across scales.

\subsection{Spectral Diversity Regularization}
\label{sec:decorrelation}

A learned operator bank can suffer from \emph{filter collapse}, where multiple components become redundant and the mixture degenerates into averaging near-identical experts.
To encourage complementary spectral behaviors, we introduce \textbf{Spectral Diversity Regularization}.
Instead of constraining embeddings directly, we regularize each component in the ranking space by enforcing diversity over preference margins, which promotes genuinely complementary experts.

\subsubsection{Spectral-Expert Ranking Margins}
For a triple $(u,i^{+},i^{-})$, we compute the preference margin under each spectral component $k\in\{1,\ldots,K\}$.
Let $\hat{\mathbf{h}}^{(k)}_{u}$ and $\tilde{\mathbf{h}}^{(k)}_{i}$ denote the component-specific representations used for scoring under component $k$.
We define the margin as
\begin{equation}
\delta^{(k)}(u,i^{+},i^{-})
=
\langle \hat{\mathbf{h}}^{(k)}_{u}, \tilde{\mathbf{h}}^{(k)}_{i^{+}} \rangle
-
\langle \hat{\mathbf{h}}^{(k)}_{u}, \tilde{\mathbf{h}}^{(k)}_{i^{-}} \rangle .
\end{equation}
For a mini-batch of size $B$, we collect margins into $\boldsymbol{\delta}^{(k)}\in\mathbb{R}^{B}$, where each entry corresponds to one sampled triple.

\subsubsection{Decorrelation in Ranking Space}
We quantify redundancy by computing an empirical correlation matrix over the batch margin vectors.
For each expert $k$, we standardize $\boldsymbol{\delta}^{(k)}$ by its batch mean and standard deviation with $\epsilon$ for stability, and define $\mathbf{C}\in\mathbb{R}^{K\times K}$ as
$\mathbf{C}_{ij}=\frac{1}{B}\bigl(\hat{\boldsymbol{\delta}}^{(i)}\bigr)^{\top}\hat{\boldsymbol{\delta}}^{(j)}$.
We penalize off-diagonal correlations while preventing a trivial collapse where all margins shrink toward zero:
\begin{equation}
\mathcal{L}_{\mathrm{div}}
=
\left\lVert \mathbf{C}-\mathbf{I} \right\rVert_{F}^{2}
+
\lambda_{\mathrm{var}}
\sum_{k=1}^{K}
\max\Bigl(0,\, \sigma_{\min}-\mathrm{std}(\boldsymbol{\delta}^{(k)})\Bigr),
\end{equation}
where $\mathbf{I}$ is the identity matrix and $\sigma_{\min}$ enforces a minimum margin variability within each expert.

%Overall, this objective encourages the operator mixture to maintain \emph{distinct spectral roles} (i.e., different frequency emphases) while preserving non-trivial ranking behavior.

\subsection{Model Optimization}
\label{sec:optimization}

We optimize TimeMM with a standard implicit-feedback objective, jointly regularized by the proposed spectral diversity term.
For each observed interaction $(u,i^{+})\in\mathcal{E}$, we sample negative items $i^{-}\sim\mathcal{N}(u)$ that user $u$ did not interact with.
We predict the user--item matching score by an inner product
$\hat{y}_{ui}=\langle \hat{\mathbf{h}}_{u}, \tilde{\mathbf{h}}_{i}\rangle$,
where $\hat{\mathbf{h}}_{u}$ is the final representations produced by spectral-aware modality routing, and $\tilde{\mathbf{h}}_{i}$ is the representations produced by time-conditioned adaptive spectral filtering.

We adopt the binary cross-entropy loss with negative sampling:
\begin{equation}
\mathcal{L}_{\mathrm{rec}}
=
-\sum_{(u,i^{+})\in\mathcal{E}} \log \sigma(\hat{y}_{ui^{+}})
-\sum_{(u,i^{-})\in\mathcal{N}} \log \bigl(1-\sigma(\hat{y}_{ui^{-}})\bigr),
\end{equation}
where $\sigma(\cdot)$ is the sigmoid function and $\mathcal{N}$ denotes the set of sampled negative pairs. The overall training objective is
\begin{equation}
\mathcal{L}
=
\mathcal{L}_{\mathrm{rec}}
+
\lambda\,\mathcal{L}_{\mathrm{div}}
+
\gamma \lVert \Theta \rVert_{2}^{2},
\end{equation}
where $\lambda$ controls its strength, and $\gamma$ is the coefficient of $\ell_{2}$ regularization over model parameters $\Theta$.

TimeMM scales linearly with the graph size because its cost is dominated by sparse message passing. With $K$ temporal operators and $L$ propagation layers, the propagation cost is $O(KL|\mathcal{E}|D)$, while the routing networks add only a lightweight $O((U+I)D)$ overhead. In terms of memory, TimeMM stores $O(ND)$ node embeddings and $K$ sets of sparse edge weights on the same topology, requiring $O(ND + K|\mathcal{E}|)$ space, and it avoids the $O(N^3)$ eigendecomposition/SVD cost of explicit spectral methods.

%%%%%%%%

\subsection{Theoretical Interpretation of Time-as-Operator}
\label{sec:theory_spectral_interpretation_concise}

We provide a brief theoretical interpretation for TimeMM.
Our goal is not to derive new theoretical claims.
Instead, we explain why \textbf{Time-as-Operator} and adaptive mixing can be viewed as a form of polynomial spectral filtering.
This perspective captures time-conditioned smoothing while avoiding explicit eigendecomposition.

\paragraph{Temporal-kernel operator bank.}
For each scale $k\in\{1,\ldots,K\}$, TimeMM applies a temporal proximity kernel (Eq.~\eqref{eq:kernel}) to reweight observed interactions and forms a weighted adjacency $\mathbf{A}^{(k)}$, inducing the symmetric normalized operator
\begin{equation}
\mathbf{S}^{(k)} \;=\; \bigl(\mathbf{D}^{(k)}\bigr)^{-1/2}\mathbf{A}^{(k)}\bigl(\mathbf{D}^{(k)}\bigr)^{-1/2},
\qquad
\mathbf{D}^{(k)}_{ii}=\sum_j \mathbf{A}^{(k)}_{ij},
\end{equation}
which yields a multi-scale operator bank $\{\mathbf{S}^{(1)},\ldots,\mathbf{S}^{(K)}\}$ sharing the same topology but different temporal weights.

\paragraph{Propagation as polynomial spectral filtering.}
With $\mathbf{X}^{(k,l+1)}=\mathbf{S}^{(k)}\mathbf{X}^{(k,l)}$, the LightGCN-style aggregation can be written as
\begin{equation}
\label{eq:concise_poly}
\mathbf{h}^{(k)} \;=\; \sum_{l=0}^{L}\bigl(\mathbf{S}^{(k)}\bigr)^{l}\mathbf{X}^{(0)} \;=\; p_L\!\bigl(\mathbf{S}^{(k)}\bigr)\mathbf{X}^{(0)},
\qquad
p_L(\lambda)=\sum_{l=0}^{L}\lambda^{l}.
\end{equation}
When $\mathbf{S}^{(k)}$ is symmetric, $\mathbf{h}^{(k)}=\mathbf{U}^{(k)}p_L(\mathbf{\Lambda}^{(k)})(\mathbf{U}^{(k)})^\top\mathbf{X}^{(0)}$, matching the standard spectral filtering form. TimeMM does not compute $\mathbf{U}^{(k)}$. It realizes $p_L(\mathbf{S}^{(k)})$ via sparse multiplications, which is both efficient and well-suited to scale-varying operators.
\emph{Intuitively}, polynomial responses such as $p_L$ behave as low-pass--like smoothing responses under common stabilized propagators (e.g., normalized operators with eigenvalues bounded in $[-1,1]$, or ``lazy'' variants that shift mass toward $[0,1]$), since repeated propagation attenuates components that vary sharply over heavily weighted edges.

\paragraph{Spectral response intuition.}
When $\mathbf{S}^{(k)}$ is symmetric, Eq.~\eqref{eq:concise_poly} implies that each spectral component along an eigenvector of $\mathbf{S}^{(k)}$ is scaled by the response $p_L(\lambda)$.
Specifically, for $\mathbf{S}^{(k)}=\mathbf{U}^{(k)}\mathbf{\Lambda}^{(k)}(\mathbf{U}^{(k)})^\top$,
\begin{equation}
\label{eq:concise_freq_response}
\mathbf{h}^{(k)}
\;=\;
\sum_{r} p_L\!\bigl(\lambda^{(k)}_{r}\bigr)\,
\mathbf{u}^{(k)}_{r}\bigl(\mathbf{u}^{(k)}_{r}\bigr)^\top \mathbf{X}^{(0)},
\qquad
p_L(\lambda)=\sum_{l=0}^{L}\lambda^l .
\end{equation}
Under common stabilized propagators where $|\lambda|\le 1$ and typically $\lambda$ concentrates near $[0,1]$, $p_L(\lambda)$ assigns larger gain to low-variation components (large $\lambda$) and comparatively suppresses high-variation ones (small or negative $\lambda$), matching a low-pass--like smoothing interpretation.

\paragraph{Effective smoothing across temporal scales.}
Each $\mathbf{S}^{(k)}$ defines a temporally weighted neighborhood geometry. Polynomial aggregation is smoothing on that geometry, and TimeMM’s multi-scale behavior arises from which temporal edges are emphasized, rather than from switching to an explicit high-pass filter. Small-scale operators emphasize recency-weighted edges and preserve local variations that would be over-smoothed by long-horizon operators, while large-scale operators aggregate over broader temporal influence and emphasize stable preference structure.

\paragraph{Time-weighted consistency and adaptive mixing.}
Let $\mathbf{L}^{(k)}=\mathbf{I}-\mathbf{S}^{(k)}$ be the normalized Laplacian. The Dirichlet energy $\mathrm{tr}(\mathbf{X}^\top\mathbf{L}^{(k)}\mathbf{X})$ measures representation inconsistency across edges weighted by $\mathbf{A}^{(k)}$, and thus encodes a time-weighted consistency principle. Concretely, for $\mathbf{X}=[\mathbf{x}_1;\ldots;\mathbf{x}_N]$,
\begin{equation}
\label{eq:concise_dirichlet}
\mathrm{tr}\!\left(\mathbf{X}^\top\mathbf{L}^{(k)}\mathbf{X}\right)
\;=\;
\frac{1}{2}\sum_{i,j}\mathbf{A}^{(k)}_{ij}\,
\left\lVert \frac{\mathbf{x}_i}{\sqrt{\mathbf{D}^{(k)}_{ii}}}-\frac{\mathbf{x}_j}{\sqrt{\mathbf{D}^{(k)}_{jj}}}\right\rVert_2^2,
\end{equation}
so large (resp.\ small) temporal scales emphasize consistency across long-horizon (resp.\ recency-weighted) edges.

Then TimeMM mixes the operator-bank outputs with temporal-state weights (Sec.~\ref{sec:scale_gating}),
\begin{equation}
\label{eq:concise_mixing}
\tilde{\mathbf{h}}(u)=\sum_{k=1}^{K} g_{u,k}\,\mathbf{h}^{(k)}(u),
\qquad
\tilde{\mathbf{h}}(i)=\sum_{k=1}^{K} g_{i,k}\,\mathbf{h}^{(k)}(i),
\end{equation}
which can be interpreted as context-conditioned selection over multiple time-weighted smoothing regimes.
Substituting Eq.~\eqref{eq:concise_poly} into Eq.~\eqref{eq:concise_mixing} shows that the fused representation is a mixture over polynomial filters $\{p_L(\mathbf{S}^{(k)})\}$:
\begin{equation}
\label{eq:concise_mixture}
\tilde{\mathbf{h}}(\cdot)
\;=\;
\sum_{k=1}^{K} g_{\cdot,k}\, p_L\!\bigl(\mathbf{S}^{(k)}\bigr)\mathbf{X}^{(0)}(\cdot),
\end{equation}
so the temporal context allocates more mass to operators whose time-weighted consistency (Eq.~\eqref{eq:concise_dirichlet}) better matches the current decision regime. Finally, TimeMM couples adaptive operator mixing with temporal state conditioned modality routing (Sec.~\ref{sec:modality_routing}), allowing the multimodal mixture to co-evolve with the selected operator regime.
Together, this provides a coherent spectral view of TimeMM without explicit decomposition.

%%%%%%%%%%%%%%%%%%%%%%%%%%%%%%%%%%%%%%%%%%%%%%%%%%%%%%%%%%%%%%%%%%%%%%

\begin{table*}[t]
\centering
\caption{
Overall recommendation performance on five datasets (four Amazon domains and one industrial offline dataset). \textbf{TimeMM} achieves the best results across all metrics, highlighting its effectiveness in capturing temporal dynamics for multimodal preference modeling.
}
\label{tab:main_results_TimeMM}
\resizebox{\textwidth}{!}{
\begin{tabular}{lccccccccccccccc}
\toprule
\textbf{Dataset} & \textbf{Metric} & BPR & LightGCN & VBPR & MMGCN & GRCN & SLMRec & LATTICE & BM3 & FREEDOM & MMIL & AlignRec & SMORE & FITMM & \textbf{TimeMM} \\
\midrule

\multirow{4}{*}{Industry}
& R@10  & 0.1032 & 0.1158 & 0.1119 & 0.1094 & 0.1162 & 0.1183 & 0.1176 & 0.0968 & 0.1043 & 0.1172 & 0.1178 & 0.1151 & \underline{0.1221} & \textbf{0.1271} \\
& R@20  & 0.1562 & 0.1675 & 0.1639 & 0.1636 & 0.1715 & 0.1687 & 0.1714 & 0.1426 & 0.1577 & 0.1721 & 0.1711 & 0.1659 & \underline{0.1763} & \textbf{0.1821} \\
& N@10  & 0.0538 & 0.0620 & 0.0590 & 0.0572 & 0.0612 & 0.0636 & 0.0633 & 0.0520 & 0.0544 & 0.0634 & 0.0630 & 0.0629 & \underline{0.0652} & \textbf{0.0686} \\
& N@20  & 0.0671 & 0.0750 & 0.0721 & 0.0709 & 0.0751 & 0.0763 & 0.0768 & 0.0636 & 0.0678 & 0.0772 & 0.0764 & 0.0747 & \underline{0.0795} & \textbf{0.0826} \\
\midrule

\multirow{4}{*}{CD}
& R@10  & 0.0424 & 0.0472 & 0.0500 & 0.0477 & 0.0493 & 0.0533 & 0.0521 & 0.0438 & 0.0585 & 0.0598 & 0.0575 & 0.0607 & \underline{0.0633} & \textbf{0.0682} \\
& R@20  & 0.0672 & 0.0765 & 0.0787 & 0.0762 & 0.0775 & 0.0836 & 0.0816 & 0.0702 & 0.0939 & 0.0952 & 0.0913 & 0.0981 & \underline{0.0986} & \textbf{0.1049} \\
& N@10  & 0.0217 & 0.0250 & 0.0252 & 0.0237 & 0.0258 & 0.0283 & 0.0275 & 0.0228 & 0.0306 & 0.0309 & 0.0295 & 0.0312 & \underline{0.0325} & \textbf{0.0356} \\
& N@20  & 0.0279 & 0.0323 & 0.0324 & 0.0308 & 0.0329 & 0.0359 & 0.0349 & 0.0295 & 0.0395 & 0.0398 & 0.0380 & 0.0405 & \underline{0.0412} & \textbf{0.0442} \\
\midrule

\multirow{4}{*}{Game}
& R@10  & 0.0463 & 0.0539 & 0.0473 & 0.0492 & 0.0510 & 0.0595 & 0.0575 & 0.0533 & 0.0561 & 0.0587 & 0.0596 & 0.0593 & \underline{0.0603} & \textbf{0.0662} \\
& R@20  & 0.0707 & 0.0809 & 0.0745 & 0.0750 & 0.0794 & 0.0913 & 0.0881 & 0.0829 & 0.0867 & 0.0886 & 0.0910 & 0.0895 & \underline{0.0914} & \textbf{0.0986} \\
& N@10  & 0.0247 & 0.0287 & 0.0242 & 0.0263 & 0.0264 & 0.0317 & 0.0307 & 0.0286 & 0.0299 & 0.0313 & 0.0320 & 0.0315 & \underline{0.0321} & \textbf{0.0354} \\
& N@20  & 0.0308 & 0.0355 & 0.0310 & 0.0328 & 0.0336 & 0.0398 & 0.0384 & 0.0360 & 0.0376 & 0.0388 & 0.0397 & 0.0391 & \underline{0.0398} & \textbf{0.0431} \\
\midrule

\multirow{4}{*}{Baby}
& R@10  & 0.0195 & 0.0215 & 0.0204 & 0.0198 & 0.0274 & 0.0314 & 0.0306 & 0.0282 & 0.0310 & 0.0332 & \underline{0.0346} & 0.0336 & 0.0328 & \textbf{0.0393} \\
& R@20  & 0.0332 & 0.0342 & 0.0339 & 0.0331 & 0.0468 & 0.0490 & 0.0511 & 0.0463 & 0.0513 & 0.0547 & \underline{0.0569} & 0.0550 & 0.0563 & \textbf{0.0602} \\
& N@10  & 0.0093 & 0.0112 & 0.0100 & 0.0097 & 0.0137 & 0.0160 & 0.0158 & 0.0142 & 0.0149 & 0.0162 & 0.0167 & \underline{0.0169} & 0.0163 & \textbf{0.0195} \\
& N@20  & 0.0127 & 0.0144 & 0.0134 & 0.0130 & 0.0185 & 0.0204 & 0.0209 & 0.0188 & 0.0201 & 0.0215 & 0.0223 & \underline{0.0223} & 0.0218 & \textbf{0.0248} \\
\midrule

\multirow{4}{*}{Software}
& R@10  & 0.1210 & 0.1212 & 0.1227 & 0.1236 & 0.1259 & 0.1274 & 0.1260 & 0.1238 & 0.1291 & 0.1280 & 0.1283 & \underline{0.1336} & 0.1328 & \textbf{0.1387} \\
& R@20  & 0.1807 & 0.1811 & 0.1884 & 0.1887 & 0.1907 & 0.1886 & 0.1866 & 0.1800 & 0.1895 & 0.1875 & 0.1890 & 0.1937 & \underline{0.1951} & \textbf{0.1995} \\
& N@10  & 0.0640 & 0.0646 & 0.0637 & 0.0652 & 0.0658 & 0.0679 & 0.0668 & 0.0657 & 0.0680 & 0.0672 & 0.0691 & \underline{0.0719} & 0.0705 & \textbf{0.0761} \\
& N@20  & 0.0791 & 0.0796 & 0.0802 & 0.0815 & 0.0821 & 0.0833 & 0.0821 & 0.0798 & 0.0832 & 0.0821 & 0.0844 & \underline{0.0870} & 0.0867 & \textbf{0.0912} \\
\bottomrule
\end{tabular}
}
\end{table*}

\section{EXPERIMENTS}
\subsection{Experimental Settings}

\paragraph{Datasets.}
We evaluate on four multimodal benchmarks constructed from the Amazon Review corpus and product metadata (\textbf{CD, Game, Baby, Software})\footnote{\url{http://jmcauley.ucsd.edu/data/amazon/}}, where each item is aligned with its associated text and image features. Additionally, we construct a real-world \textbf{Industrial offline dataset} to assess robustness under realistic distributions. Dataset statistics are summarized in Table~\ref{tab:datasets_tatistics}.

\paragraph{Baselines.}
We compare against two interaction-only baselines (BPR~\cite{rendle2012bpr}, LightGCN~\cite{he2020lightgcn}) and a broad set of multimodal recommenders, including VBPR~\cite{he2016vbpr}, MMGCN~\cite{wei2019mmgcn}, GRCN~\cite{wei2020graph}, DualGNN~\cite{wang2021dualgnn}, SLMRec~\cite{tao2022self}, LATTICE~\cite{zhang2021mining}, BM3~\cite{zhou2023bootstrap}, FREEDOM~\cite{zhou2023tale}, DiffMM~\cite{jiang2024diffmm}, MMIL~\cite{yang2024multimodal}, AlignRec~\cite{liu2024alignrec}, and spectrum-aware methods SMORE~\cite{ong2025spectrum} and FITMM~\cite{yang2025fitmm}.

\paragraph{Evaluation and Implementation.}
We adopt all-ranking evaluation with a leave-one-out split and report Recall@K and NDCG@K for $K\in\{10,20\}$. For each user, interactions are first sorted in chronological order. The most recent interaction is used for testing, the second most recent one is used for validation, and all earlier interactions are used for training. This user-level temporal split is applied consistently to all datasets, including the Industrial offline dataset, which prevents future interactions from appearing in the training set. We further ensure that all time-dependent statistics and temporal context states in TimeMM are constructed using training interactions only, without accessing validation or test interactions. Therefore, no future timestamp information is used when building temporal states, temporal kernels, or operator mixtures during training. We use standard negative sampling during training. All models are implemented in PyTorch with MMRec~\cite{zhou2023mmrec}. Unless stated otherwise, we use embedding size 64, batch size 2048, and early stopping on Recall@20. We train the model with Adam~\cite{kingma2014adam}, selecting the learning rate from $\{0.0001, 0.0005, 0.001, 0.005\}$. For hyperparameters, we tune the diversity loss coefficient $\lambda$ and the regularization coefficient $\gamma$ over $\{0.0001, 0.001, 0.01, 0.1, 1.0\}$. We also vary the number of scales $K$ in $\{1,2,3,4,5\}$ to study the effect of multi-scale decomposition.

\begin{table}[t]
  \caption{Statistics of the experimental datasets constructed from Amazon raw data and real-world industrial data.}
  \label{tab:datasets_tatistics}
  \centering
  \begin{tabular}{lcccc}
    \toprule
    Dataset & \#Users & \#Items & \#Interactions & Sparsity \\
    \midrule
    Industry & 59,989  & 12,281  & 1,023,597   & 99.861\% \\
    CD       & 22,021  & 19,971  & 337,360   & 99.923\% \\
    Software & 29,798  & 6,342   & 425,222   & 99.775\% \\
    Baby     & 149,732 & 35,742  & 1,166,408 & 99.978\% \\
    Game    & 94,716  & 25,596  & 755,431   & 99.969\% \\
    \bottomrule
  \end{tabular}
\end{table}

\subsection{RQ1: Overall Performance}
Table~\ref{tab:main_results_TimeMM} summarizes the overall performance of different baselines on five datasets. 
Our \textbf{TimeMM} achieves the best results across all datasets and metrics, demonstrating strong generalization from academic benchmarks to a realistic industrial distribution. 
The gains are consistent and practically meaningful: relative to the strongest baseline on each dataset, TimeMM typically improves Recall and NDCG by about $3\%$--$10\%$. For example, on Amazon CD it boosts R@10 from 0.0633 to 0.0682 ($+7.7\%$) and N@10 from 0.0325 to 0.0356 ($+9.5\%$). 
Compared with general multimodal graph recommenders that enhance fusion or cross-view consistency on a static snapshot such as AlignRec and FREEDOM, TimeMM is consistently stronger, indicating that temporal non-stationarity cannot be addressed by static multimodal refinements alone. 
It also outperforms frequency-based multimodal recommenders such as SMORE and FITMM, whose static spectral decompositions inherit a stationarity assumption and struggle to reflect time-varying spectral shifts. 
In contrast, TimeMM realizes frequency-aware modeling implicitly via Time-as-Operator by constructing a temporal-kernel operator bank with adaptive operator mixing and spectral-aware modality routing.
Meanwhile, the spectral diversity regularizer prevents collapse and promotes complementary operators, yielding consistent gains even over the strongest spectral baselines.
%These results support that modeling time as a first-class operator yields systematic benefits beyond architectural tweaks.

%A closer comparison with representative baselines further clarifies why TimeMM is consistently superior. Against strong general multimodal graph recommenders, the gains highlight the necessity of explicitly modeling temporal dynamics rather than enhancing fusion or denoising on a static interaction snapshot. For instance, AlignRec strengthens cross-view alignment but remains temporally agnostic, TimeMM still achieves clear improvements. Similarly, FREEDOM improves robustness via denoised relations, yet cannot distinguish outdated evidence from emerging intent, whereas TimeMM benefits from treating time as an operator to construct scale-specific propagation operators and from adaptive scale filtering to select an effective temporal scope conditioned on temporal states. Moreover, TimeMM also outperforms frequency-based multimodal recommenders, including SMORE and FITMM. Although these methods attempt to learn multi-band representations, they are typically built upon static graph spectral modeling and thus implicitly assume stationarity, which cannot capture time-varying spectral shifts. In contrast, TimeMM realizes frequency-aware behavior through dynamic, time-parameterized operators and further enforces complementary scale behaviors via diversity regularization, leading to consistent gains even over the strongest spectral baseline.

\subsection{RQ2: Ablation Study}
\begin{figure}[h]
  \centering
  \includegraphics[width=\linewidth]{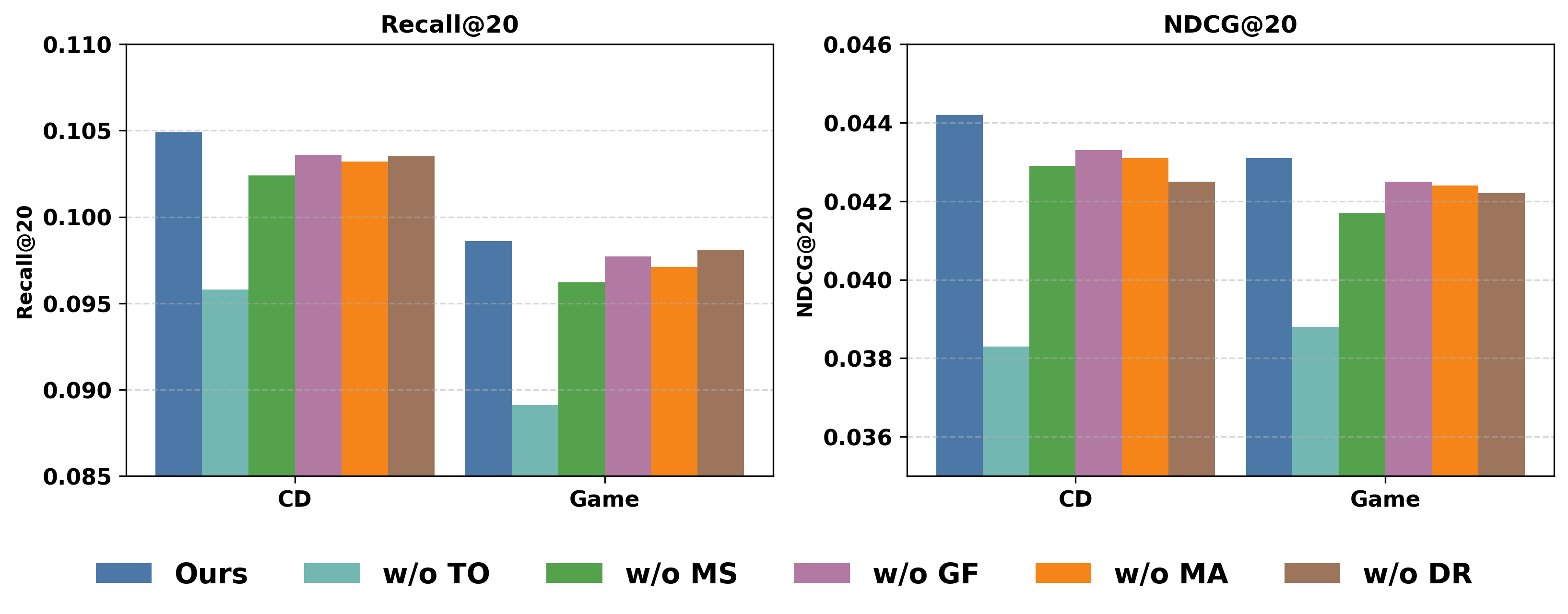}
  \caption{
Ablation study of \textbf{TimeMM} on three datasets. Each variant removes one key component: \textbf{TO} (Time-as-Operator), \textbf{MS} (multi-scale propagation), \textbf{GF} (gated filtering over scales), \textbf{MA} (spectral-aware modality routing), and \textbf{DR} (scale diversity regularization). We report Recall@20 (R20) and NDCG@20 (N20).
}
  \label{fig:exp_ablation_TimeMM}
\end{figure}

We conduct ablation studies to quantify the contribution of each key module in TimeMM. Figure~\ref{fig:exp_ablation_TimeMM} summarizes five variants that remove one component at a time: \textbf{w/o TO} removes \textbf{Time-as-Operator} and thus eliminates temporal-kernel-induced operator modulation; \textbf{w/o MS} disables the multiple spectral operators and reduces the model to a single smoothing profile; \textbf{w/o GF} removes adaptive spectral filtering with operator mixing, weakening context-conditioned selection; \textbf{w/o MA} removes spectral-aware modality routing; and \textbf{w/o DR} removes the spectral diversity regularizer. Among all variants, \textit{w/o TO} leads to the largest degradation, indicating that injecting temporal dynamics into the propagation operator is the primary driver of TimeMM. For instance, on CD, removing TO reduces R20 from 0.1049 to 0.0918 and N20 from 0.0442 to 0.0383, reflecting a pronounced loss in both retrieval accuracy and ranking quality. In contrast, \textit{w/o MS} and \textit{w/o GF} yield smaller yet consistent decreases, suggesting complementary benefits beyond a single fixed spectral response. The operator bank provides diverse spectral views of preference signals, while adaptive mixing selects an effective response conditioned on temporal context. Removing \textit{MA} also causes consistent declines, supporting the motivation that modality contributions should be routed according to temporal context rather than remain fixed. Finally, \textit{w/o DR} shows a modest but non-negligible drop, implying that ranking-space spectral diversity may lightly help prevent expert collapse and improve the effectiveness of adaptive operator mixing. 
%Together, these results validate that TimeMM’s gains arise from a coherent combination of time-conditioned operators, adaptive spectral filtering, modality routing, and spectral diversity regularization.

\subsection{RQ3: How Do Timestamps Affect Performance?}
\label{sec:rq4_time_perturbation}

To directly test whether TimeMM benefits from meaningful temporal structure rather than incidental correlations, we perform a controlled timestamp perturbation study on CD and Game.
We keep the user--item interaction identities unchanged and only manipulate timestamps, isolating the contribution of time from the static interaction topology.
We consider four settings: (1) \textbf{Original} uses the raw timestamps; (2) \textbf{Shuffle} randomly permutes timestamps within each user, breaking temporal order and recency; (3) \textbf{Constant} maps all timestamps to the same value, removing temporal variation; and (4) \textbf{Noise} adds random jitter to timestamps, preserving the rough order while perturbing temporal distances.

Table~\ref{tab:time_perturbation} reports the results. Across both datasets, TimeMM is most sensitive to destroying relative temporal order.
\textit{Shuffle} causes the largest drop (e.g., on CD, R@20 decreases from 0.1049 to 0.0869 and N@20 from 0.0442 to 0.0353).
This is expected under \textbf{Time-as-Operator}: timestamps are mapped to temporal-kernel weights that define a bank of propagation operators with distinct smoothing profiles.
When order is randomized, the induced operators no longer reflect meaningful recency structure, and adaptive spectral filtering cannot recover a coherent effective response.
\textit{Constant} also degrades performance, but is generally less harmful than \textit{Shuffle}, indicating that preserving a consistent recency relation is more critical than retaining absolute time values.
Finally, \textit{Noise} yields the smallest degradation and remains close to \textit{Original}, suggesting that TimeMM is robust to moderate timestamp noise common in real logging systems while still exploiting order-consistent temporal cues to some extent.

\begin{table}[t]
\centering
\small
\setlength{\tabcolsep}{5pt}
\begin{tabular}{llcccc}
\toprule
\textbf{Dataset} & \textbf{Metric} & \textbf{Origin (Base)} & \textbf{Shuffle} & \textbf{Constant} & \textbf{Noise} \\
\midrule
\multirow{4}{*}{CD}
& R@10 & \textbf{0.0682} & 0.0561 & 0.0604 & 0.0655 \\
& R@20 & \textbf{0.1049} & 0.0869 & 0.0961 & 0.1018 \\
& N@10 & \textbf{0.0356} & 0.0279 & 0.0313 & 0.0336 \\
& N@20 & \textbf{0.0442} & 0.0353 & 0.0401 & 0.0424 \\
\midrule
\multirow{4}{*}{Game}
& R@10 & \textbf{0.0662} & 0.0532 & 0.0538 & 0.0628 \\
& R@20 & \textbf{0.0986} & 0.0806 & 0.0813 & 0.0927 \\
& N@10 & \textbf{0.0354} & 0.0284 & 0.0295 & 0.0345 \\
& N@20 & \textbf{0.0431} & 0.0341 & 0.0352 & 0.0410 \\
\bottomrule
\end{tabular}
\caption{Timestamp perturbation study on CD and Game. We keep user--item interactions unchanged and only perturb timestamps. \textbf{Original} uses raw timestamps; \textbf{Shuffle} permutes timestamps within each user (breaking temporal recency); \textbf{Constant} maps all timestamps to a single value (removing temporal variation); \textbf{Noise} adds random jitter to timestamps (introducing measurement errors while roughly preserving order).}
\label{tab:time_perturbation}
\end{table}

\subsection{RQ4: Who Benefits Most? Performance by User History Span}
\label{sec:rq6_span_bucket}

A core motivation of TimeMM is that time-conditioned spectral modeling becomes most valuable when temporal non-stationarity is observable. We measure each user's history span as $\Delta_u=t_u^{\max}-t_u^{\min}$ and partition users into three quantile buckets: short-span (B1), mid-span (B2), and long-span (B3). We then compare TimeMM with a strong spectral baseline, SMORE, under the same protocol. Table~\ref{tab:span_bucket} shows a consistent pattern: \textbf{TimeMM's advantage increases with user history span}, with the largest gains concentrated in B3. On CD, TimeMM improves SMORE on B3 from 0.0739 to 0.0840 in R@10, and from 0.1180 to 0.1265 in R@20. On Game, B3 improves from 0.0782 to 0.0884 in R@10, and from 0.1202 to 0.1289 in R@20. This span-amplified improvement aligns with our Time-as-Operator design: Long-span users contain richer temporal structure, allowing the temporal-kernel operator bank to induce more informative smoothing profiles and enabling adaptive operator mixing and modality routing to select an effective response per prediction. In contrast, static spectral filtering on a snapshot struggles to reflect time-varying spectral shifts that become more pronounced over longer horizons. Gains are smaller for B1 and become more visible in B2 before peaking in B3, which is expected because limited temporal coverage reduces the identifiability of long-horizon drift. Importantly, TimeMM remains competitive across all buckets rather than trading off regimes, indicating that temporal operators provide benefits when temporal signal is rich without introducing instability for temporally sparse users.

\begin{table*}[t]
\centering
\small
\setlength{\tabcolsep}{3.5pt}
\begin{tabular}{llcccccccccccc}
\toprule
\textbf{Dataset} & \textbf{Metric} &
\multicolumn{3}{c}{\textbf{All-Span}} &
\multicolumn{3}{c}{\textbf{Short-Span (B1)}} &
\multicolumn{3}{c}{\textbf{Mid-Span (B2)}} &
\multicolumn{3}{c}{\textbf{Long-Span (B3)}} \\
\cmidrule(lr){3-5}\cmidrule(lr){6-8}\cmidrule(lr){9-11}\cmidrule(lr){12-14}
& &
\textbf{SMORE} & \textbf{Ours} & \textbf{$\Delta$} &
\textbf{SMORE} & \textbf{Ours} & \textbf{$\Delta$} &
\textbf{SMORE} & \textbf{Ours} & \textbf{$\Delta$} &
\textbf{SMORE} & \textbf{Ours} & \textbf{$\Delta$} \\
\midrule
\multirow{4}{*}{\textbf{CD}}
& R@10 & 0.0607 & 0.0682 & +0.75\% & 0.0465 & 0.0478 & +0.13\% & 0.0616 & 0.0722 & +1.06\% & 0.0739 & 0.0840 & +1.01\% \\
& R@20 & 0.0981 & 0.1049 & +0.68\% & 0.0751 & 0.0787 & +0.36\% & 0.1014 & 0.1098 & +0.84\% & 0.1180 & 0.1265 & +0.85\% \\
& N@10 & 0.0312 & 0.0356 & +0.44\% & 0.0223 & 0.0246 & +0.23\% & 0.0322 & 0.0377 & +0.55\% & 0.0389 & 0.0443 & +0.54\% \\
& N@20 & 0.0405 & 0.0442 & +0.37\% & 0.0301 & 0.0317 & +0.16\% & 0.0415 & 0.0466 & +0.51\% & 0.0502 & 0.0545 & +0.43\% \\
\midrule
\multirow{4}{*}{\textbf{Game}}
& R@10 & 0.0593 & 0.0662 & +0.69\% & 0.0339 & 0.0391 & +0.52\% & 0.0656 & 0.0712 & +0.56\% & 0.0782 & 0.0884 & +1.02\% \\
& R@20 & 0.0895 & 0.0986 & +0.91\% & 0.0497 & 0.0623 & +1.26\% & 0.0986 & 0.1045 & +0.59\% & 0.1202 & 0.1289 & +0.87\% \\
& N@10 & 0.0315 & 0.0354 & +0.39\% & 0.0181 & 0.0204 & +0.23\% & 0.0347 & 0.0386 & +0.39\% & 0.0417 & 0.0476 & +0.59\% \\
& N@20 & 0.0391 & 0.0431 & +0.40\% & 0.0218 & 0.0253 & +0.35\% & 0.0431 & 0.0467 & +0.36\% & 0.0523 & 0.0575 & +0.52\% \\
\bottomrule
\end{tabular}
\caption{Performance by user temporal-span buckets. Users are partitioned into three quantile buckets based on $\Delta_u=t_u^{\max}-t_u^{\min}$ (B1: short-span, B2: mid-span, B3: long-span). $\Delta$ is reported as an absolute gain in percentage-point style.}
\label{tab:span_bucket}
\end{table*}

\iffalse
\begin{table*}[t]
\centering
\small
\setlength{\tabcolsep}{4pt}
\begin{tabular}{llcccccccc}
\toprule
& & \multicolumn{4}{c}{\textbf{CD}} & \multicolumn{4}{c}{\textbf{Game}} \\
\cmidrule(lr){3-6}\cmidrule(lr){7-10}
\textbf{Model} & \textbf{Metric} &
\textbf{All} & \textbf{Short-Span (B1)} & \textbf{Mid-Span (B2)} & \textbf{Long-Span (B3)} &
\textbf{All} & \textbf{Short-Span (B1)} & \textbf{Mid-Span (B2)} & \textbf{Long-Span (B3)} \\
\midrule
\multirow{4}{*}{SMORE}
& R@10 & 0.0607 & 0.0465 & 0.0616 & 0.0739 & 0.0593 & 0.0339 & 0.0656 & 0.0782 \\
& R@20 & 0.0981 & 0.0751 & 0.1014 & 0.1180 & 0.0895 & 0.0497 & 0.0986 & 0.1202 \\
& N@10 & 0.0312 & 0.0223 & 0.0322 & 0.0389 & 0.0315 & 0.0181 & 0.0347 & 0.0417 \\
& N@20 & 0.0405 & 0.0301 & 0.0415 & 0.0502 & 0.0391 & 0.0218 & 0.0431 & 0.0523 \\
\midrule
\multirow{4}{*}{Ours}
& R@10 & 0.0682 & 0.0478 & 0.0722 & 0.0840 & 0.0662 & 0.0391 & 0.0712 & 0.0884 \\
& R@20 & 0.1049 & 0.0787 & 0.1098 & 0.1265 & 0.0986 & 0.0623 & 0.1045 & 0.1289 \\
& N@10 & 0.0356 & 0.0246 & 0.0377 & 0.0443 & 0.0354 & 0.0204 & 0.0386 & 0.0476 \\
& N@20 & 0.0442 & 0.0317 & 0.0466 & 0.0545 & 0.0431 & 0.0253 & 0.0467 & 0.0575 \\
\bottomrule
\end{tabular}
\caption{Performance by user temporal-span buckets. Users are partitioned into three quantile buckets based on $\Delta_u=t_u^{\max}-t_u^{\min}$ (Bucket 1: short-span, Bucket 2: medium-span, Bucket 3: long-span). ``All'' reports the overall performance without bucketing.}
\label{tab:span_bucket}
\end{table*}
\fi

\subsection{RQ5: Do Learned Operators Follow a Consistent Short-to-long Ordering?}
\label{sec:rq6_energy_decay}
% 有点复杂、抽象

We examine whether TimeMM’s temporal-kernel operator bank learns a meaningful short-to-long ordering, rather than collapsing into redundant operators. We instantiate TimeMM with $K=3$ operators and bucket users into \textit{short-span}, \textit{mid-span}, and \textit{long-span} groups by their history span $\Delta_u=t_u^{\max}-t_u^{\min}$. All statistics are computed on held-out user--positive-item pairs. As a scale-comparable signature of smoothing, we measure the user--positive-item discrepancy under operator $k$ as $E_k(u,i)=\lVert \mathbf{z}^{(k)}_u-\mathbf{z}^{(k)}_i\rVert_2^2/d$ (smaller $E_k$ indicates stronger smoothing), and summarize decay via ratios $r_{12}=E_2/E_1$ and $r_{13}=E_3/E_1$, equivalently drops $g_{12}=1-r_{12}$ and $g_{13}=1-r_{13}$. The key expectation is a monotone short-to-long decay $E_1 \ge E_2 \ge E_3$, i.e., $r_{12},r_{13}<1$ or $g_{12},g_{13}>0$.

\begin{figure}[h]
  \centering
  \includegraphics[width=\linewidth]{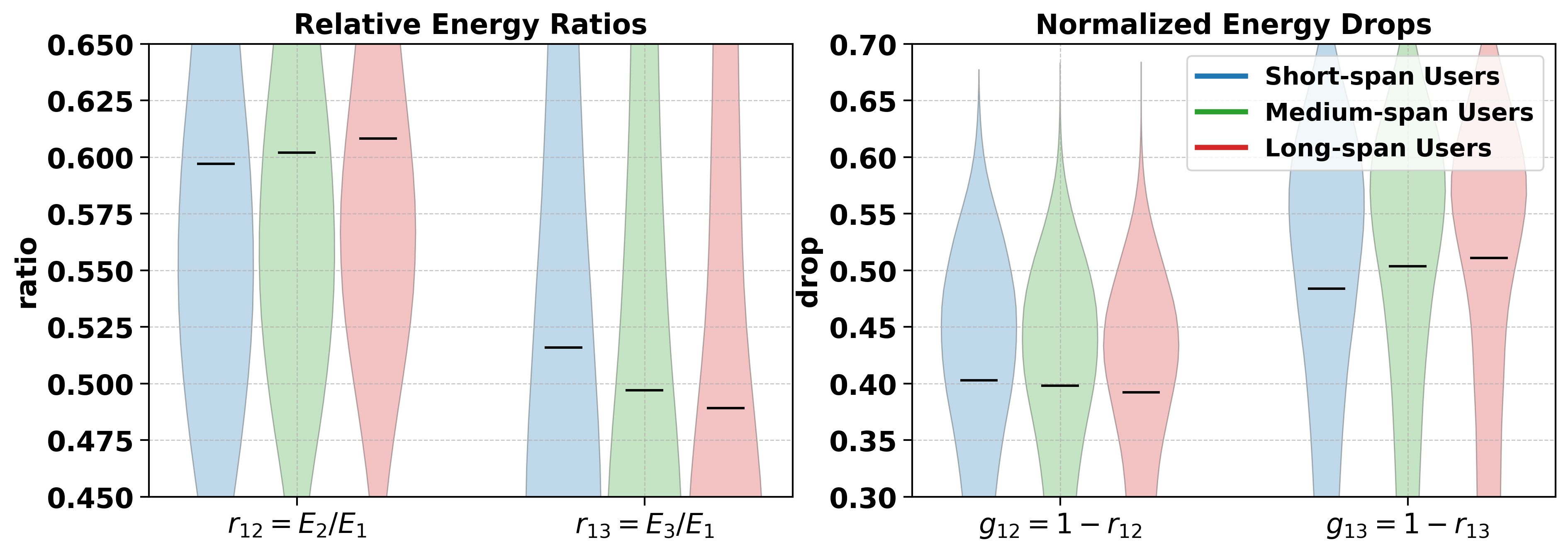}
  \caption{
Energy-decay signatures of TimeMM’s learned multi-scale operators across user temporal-span groups. The stronger separation for $r_{13}/g_{13}$ suggests that long-horizon smoothing varies systematically across user regimes.
}
  \label{fig:energy_decay_shape}
\end{figure}

Figure~\ref{fig:energy_decay_shape} confirms consistent energy decay from operator 1 to 2 and to 3 across all user groups: both $r_{12}$ and $r_{13}$ fall reliably below $1$, indicating progressively stronger attenuation rather than replica behavior. The separation across user groups is more pronounced for the long-range comparison ($r_{13}$/$g_{13}$) than for the adjacent comparison ($r_{12}$/$g_{12}$), suggesting that the strongest smoothing is most sensitive to a user’s temporal regime. This is further supported by a sample-level monotonicity check of $E_1 \ge E_2 \ge E_3$: the monotonic rate increases from 0.8637 (short-span) to 0.8901 (mid-span) and 0.9050 (long-span), while the average violation magnitude decreases from $3.38\times10^{-4}$ to $2.37\times10^{-4}$ and $1.92\times10^{-4}$. Overall, TimeMM learns an interpretable and consistently ordered operator hierarchy, with the ordering expressed most cleanly for long-span users.

\iffalse
\begin{table}[t] \centering \small \setlength{\tabcolsep}{6pt} \begin{tabular}{lccc} \toprule \textbf{Bucket} & \textbf{\#Samples} & \textbf{Monotonic rate} & \textbf{Violation mean} \\ \midrule bucket1 & 96{,}430 & 0.8637 & 3.38$\times 10^{-4}$ \\ bucket2 & 99{,}221 & 0.8901 & 2.37$\times 10^{-4}$ \\ bucket3 & 97{,}667 & 0.9050 & 1.92$\times 10^{-4}$ \\ \bottomrule \end{tabular} \caption{Sample-level energy ordering consistency across user temporal-span buckets.} \label{tab:energy_ordering_stats} \end{table}
\fi

\subsection{RQ6: User-level Operator Mixing Patterns}
\label{sec:rq7_scale_filtering}

\begin{figure}[h]
  \centering
  \includegraphics[width=\linewidth]{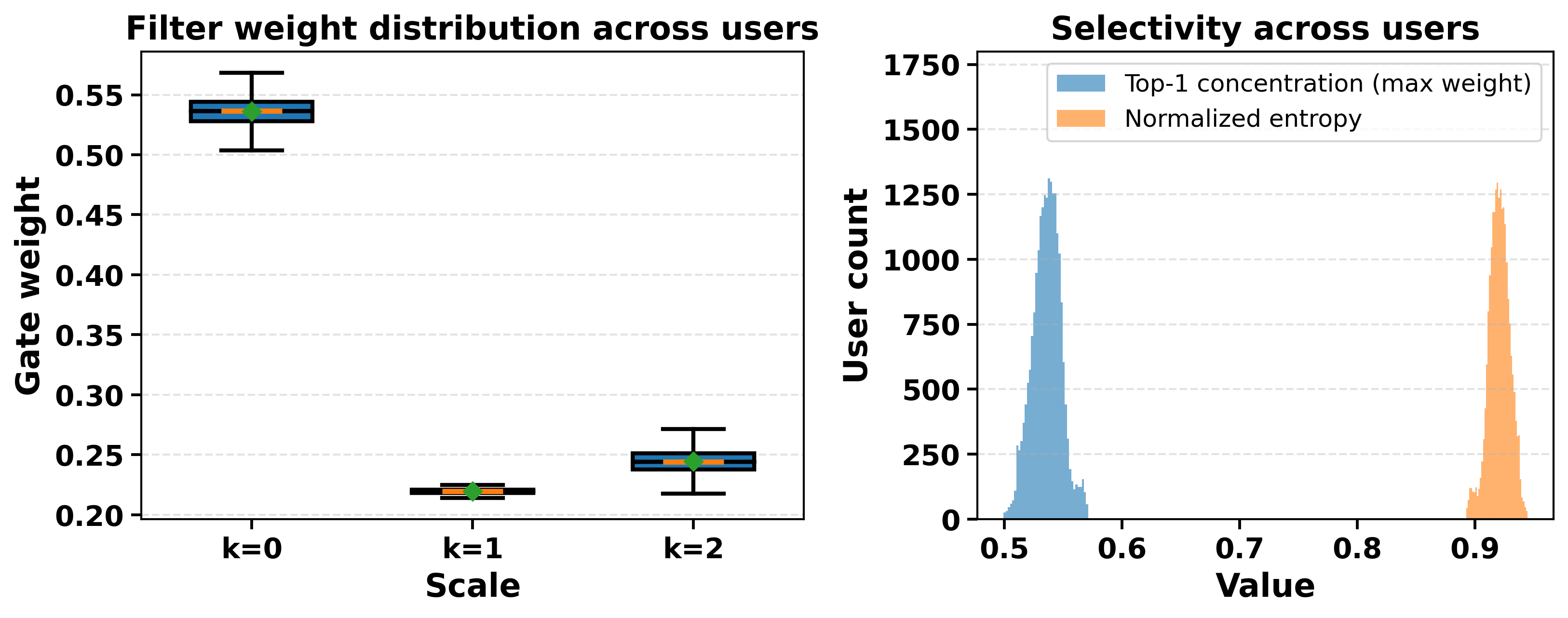}
  \caption{User-level distribution of TimeMM’s operator-mixing weights. \textbf{Left:} boxplots of fusion weights over short-, mid-, and long-horizon operators, showing a structured rather than uniform allocation. \textbf{Right:} selectivity diagnostics, indicating soft yet selective reliance on an effective temporal scope instead of uniform averaging or one-hot routing.}
  \label{fig:scale_filtering}
\end{figure}

Figure~\ref{fig:scale_filtering} characterizes TimeMM’s operator mixing at the user level. Since smaller operators emphasize recency while larger operators induce stronger long-horizon attenuation, the learned fusion weights reveal how each user balances short-, mid-, and long-range evidence when aggregating the operator-bank outputs. The left panel in Figure~\ref{fig:scale_filtering} shows that the learned weighting is \textbf{structured rather than uniform}. Across users, most mass is assigned to the shortest horizon with a median around $0.54$, indicating that recency-sensitive information is broadly predictive, while the longest-horizon operator retains a non-trivial share around $0.24$, suggesting that stable preference structure consistently benefits from long-range evidence. The intermediate horizon concentrates tightly around $0.22$ with the smallest dispersion, implying that mid-range signals behave like a relatively shared prior across users, whereas user heterogeneity is expressed primarily through how mass shifts between the shortest and the longest horizons. The right panel further indicates that the learned weighting is \textbf{selective but not degenerate}. The Top-1 concentration peaks around $0.53$--$0.56$, well above the uniform level of $1/3$ for $K{=}3$, suggesting that most users exhibit a dominant effective horizon, yet the normalized entropy remains high (around $0.90$--$0.94$), showing that the mixture does not collapse to near one-hot routing and continues to preserve secondary horizons. Together, these diagnostics support that TimeMM actively leverages the operator bank as a user-adaptive temporal filter, rather than uniformly averaging scales or relying on hard routing.

%%%%%%%%%%%%%%%
\subsection{RQ7: Modality Mixture Shifts with User Temporal Span}
\label{sec:rq8_modality_vs_span}

\begin{figure}[h]
  \centering
  \includegraphics[width=\linewidth]{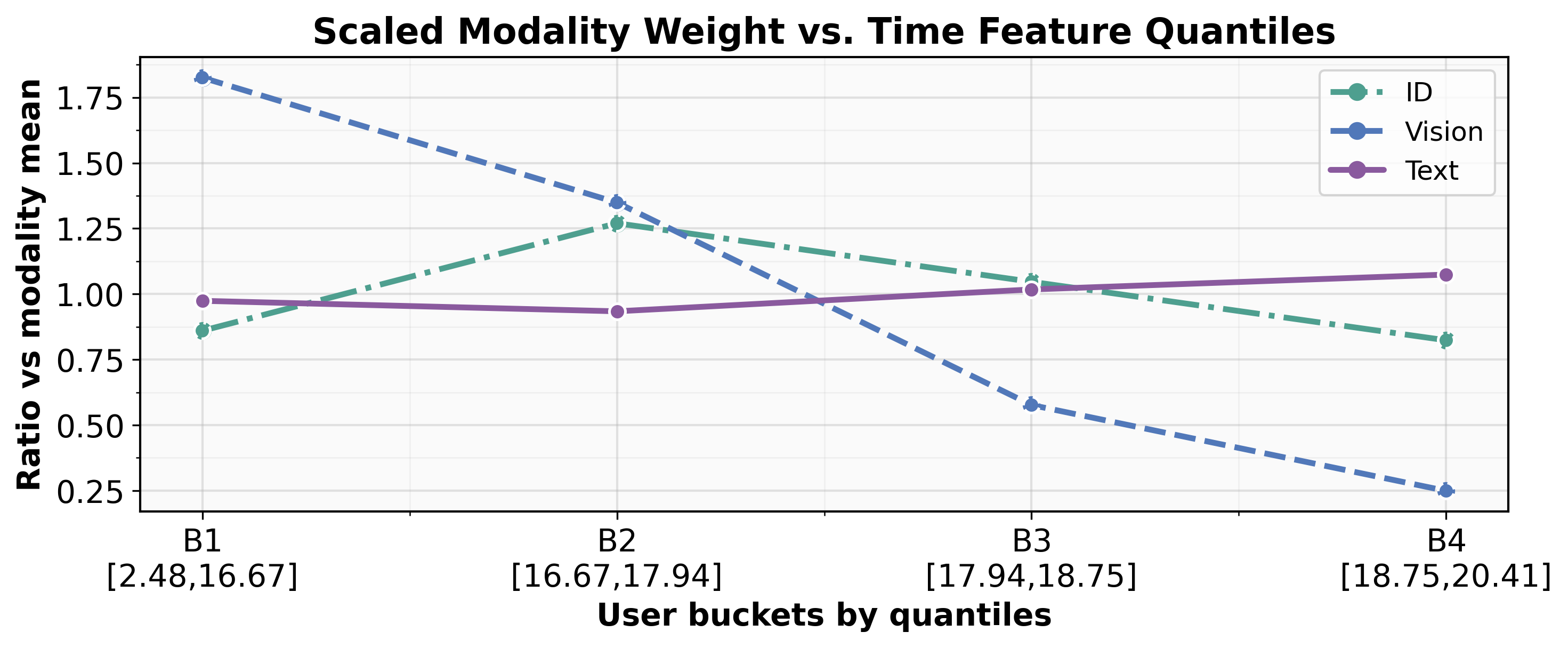}
  \caption{Span-conditioned modality mixtures. Users are partitioned into quartiles by $\log(\mathrm{span}_u)$, where $\mathrm{span}_u=t^{(u)}_{\max}-t^{(u)}_{\min}$ is computed from each user’s training interactions. For each bucket, we report the mean softmax-normalized modality weights over the ID, vision, and text channels, illustrating how the learned multimodal mixture varies with users’ span.}
  \label{fig:mod_weight_vs_log_span}
\end{figure}

We investigate whether TimeMM’s spectral-aware modality routing changes systematically with users’ temporal coverage. For each user, we compute the training span $\mathrm{span}_u=t^{(u)}_{\max}-t^{(u)}_{\min}$, bucket users into quartiles by $\log(\mathrm{span}_u)$, and summarize routing behavior per bucket. Since the absolute modality weights can differ in scale, we report the bucket-wise mean of the \emph{relative} modality weight, defined as the ratio between the bucket mean and the global mean of each modality. Figure~\ref{fig:mod_weight_vs_log_span} reveals clear span-conditioned trends. The \textbf{vision} channel decreases monotonically from short-span to long-span users, indicating that visual cues are emphasized more when user histories are temporally compact. In contrast, the \textbf{text} channel shows a mild upward trend as span increases, suggesting that semantic signals become relatively more relied upon as longer-horizon evidence accumulates. The \textbf{ID} channel is comparatively stable, exhibiting only moderate fluctuations across buckets. 
These patterns are consistent with the intuition that short-span users rely more on fast-changing signals, whereas long-span users favor stable text-based semantics. Since TimeMM does not impose any monotonic constraint between $\log(\mathrm{span}_u)$ and routing weights, these trends reflect empirical regularities learned from data rather than a hard-coded artifact.

%%%%%%%%%%%%%%%%%%%%%%%%%%%%%%%%%%%%%%%%%%%%%%%%%%%%%%%%%%%%%%%%%%%%%%
\section{Conclusion}
\label{sec:conclusion}
In this paper, we revisit multimodal recommendation from a time-conditioned spectral perspective and propose TimeMM, a scalable framework based on the Time-as-Operator paradigm. By converting timestamps into a temporal-kernel operator bank and performing adaptive operator mixing with spectral-aware modality routing, TimeMM models non-stationary preference evolution on the user--item graph without relying on expensive explicit eigendecomposition. Extensive experiments show consistent improvements over strong multimodal and spectral baselines, while retaining linear-time scalability in large-scale dynamic settings. Looking forward, we hope to extend Time-as-Operator beyond recommendation to broader dynamic graph learning problems, and explore LLM-based semantic supervision to make operator selection and preference evolution more interpretable.

\balance
\bibliographystyle{ACM-Reference-Format}
\bibliography{sample-base}

\end{document}